\RequirePackage{fix-cm}
\documentclass{svjour3}                     % onecolumn (standard format)
\smartqed  % flush right qed marks, e.g. at end of proof
\usepackage{graphicx}
\usepackage[numbers]{natbib}
\usepackage{hyperref}

\begin{document}

\title{Particle streak velocimetry using Ensemble Convolutional Neural Networks}
\titlerunning{Streaks imaging with CNN}        % if too long for running head

\author{Alexander V. Grayver         \and
        Jerome Noir
}

\institute{A. V. Grayver \at
              Institute of Geophysics, ETH Zurich \\
              Sonneggstrasse 5 \\
							8092 Zurich, Switzerland \\
              Tel.: +41-44-6333154\\
              \email{agrayver@erdw.ethz.ch}
              \and
           J. Noir \at
					    Institute of Geophysics, ETH Zurich \\
              Sonneggstrasse 5 \\
							8092 Zurich, Switzerland \\
              Tel.: +41-44-6337593\\
              \email{jerome.noir@erdw.ethz.ch}
}

\date{Received: - / Accepted: -}

\maketitle

\begin{abstract}
This study reports an approach and presents its open-source implementation for quantitative analysis of experimental flows using streak images and Convolutional Neural Networks (CNN). The latter are applied to retrieve a length and an angle from streaks, which can be used to deduce kinetic energy and directionality (up to 180$^{\circ}$ ambiguity) of an imaged flow. We developed a quick method for generating essentially unlimited number of training and validation images, which enabled efficient training. Additionally, we show how to apply an ensemble of CNNs to derive a formal uncertainty on the estimated quantities. The approach is validated on the numerical simulation of a convenctive turbulent flow and applied to a longitutidal libration flow experiment.
\end{abstract}

\section{Introduction}
\label{sec:intro}

Particle Image Velocimetry (PIV) is arguably the most widely used technique to quantitatively study experimental flows \cite{westerweel1997fundamentals, raffel2018particle}. A common work-flow would consist of seeding a flow with luminescent particles and taking pairs of pictures with a known short time separation to capture an instantaneous flow state. By splitting a pair of images into (possibly overlapping) windows and cross-correlating between them allows one to infer the direction and magnitude of the flow. With modern computers and digital cameras, PIV has experienced wide adoption in the academic and industrial domains \cite{tropea2007springer}.
For cross-correlation to work properly, one has to ensure that the exposure time is short enough such that particles do not move more than a few pixels and the two images have clearly identifiable correlations \cite{westerweel1997fundamentals}. Violating this condition, for instance because of an insufficient laser intensity or a too fast flow, results in so called streaks - traces of particles.

Streaks in the PIV images are commonly considered an experimental failure since they render cross-correlation techniques less efficient or even inapplicable. When flow velocity imposes constraints for which the camera and light source at hand cannot capture instantaneous particle positions, one either has to use a higher speed camera or/and a stronger light source. Both of these solutions quickly hit financial and safety constraints, which often cannot be overcome in academia. Therefore, there exists a need for processing techniques.
Although not suitable for conventional cross-correlation methods, streak images do carry information about the flow. Whereas an image with streaks does not contain information about direction of the underlying flow, information on the mean velocity magnitude and azimuth can be inferred from streaks themselves. To the best of authors' knowledge, there are no reliable algorithms to recover quantitative information from streaks images, which motivated us to design a method for extracting information about the flow from streaks. 

To this end, we applied an ensemble of Convolutional Neural Networks (CNN) trained on streak images to draw a statistical prediction of the displacement magnitude and corresponding azimuth. The potential of CNNs has long been recognized for applications related to the face and handwriting recognition \cite{lawrence1997face, simard2003best}, although their power has been fully discovered only recently when deeper (that is, with more layers) networks became feasible to train within reasonable amount of time, mostly due to the emergence of affordable Graphic Processor Units (GPUs) \cite{krizhevsky2012imagenet, karpathy2014large}. An interested reader is referred to a recent overview of the deep learning with important development milestones listed \cite{lecun2015deep}.

Before we proceed to describing methodology and results, it is worth to mention that a crucial ingredient to a success of CNNs is a suitably large training set, which a CNN is supposed to learn from without facing an over-fitting. From this perspective, streak analysis appears almost an ideal problem. Similar to PIV, we make a reasonable assumption that within a sufficiently small interrogation window velocity exhibits simple functional form (for instance, constant or linear function). This allows us to quickly generate an unlimited number of training images. Adding variability in number of streaks per window, their thickness and intensity will mimic a realistic experimental scenario sufficiently well. Therefore, we can exhaustively sample the parameter space and feed the network with as many sample as needed to attain an acceptable accuracy. The specific numbers and parameters will be discussed in the corresponding sections below.

\section{Methods}
\label{sec:methodology}

\subsection{Problem setup}

\begin{figure}
	\includegraphics[width=\textwidth]{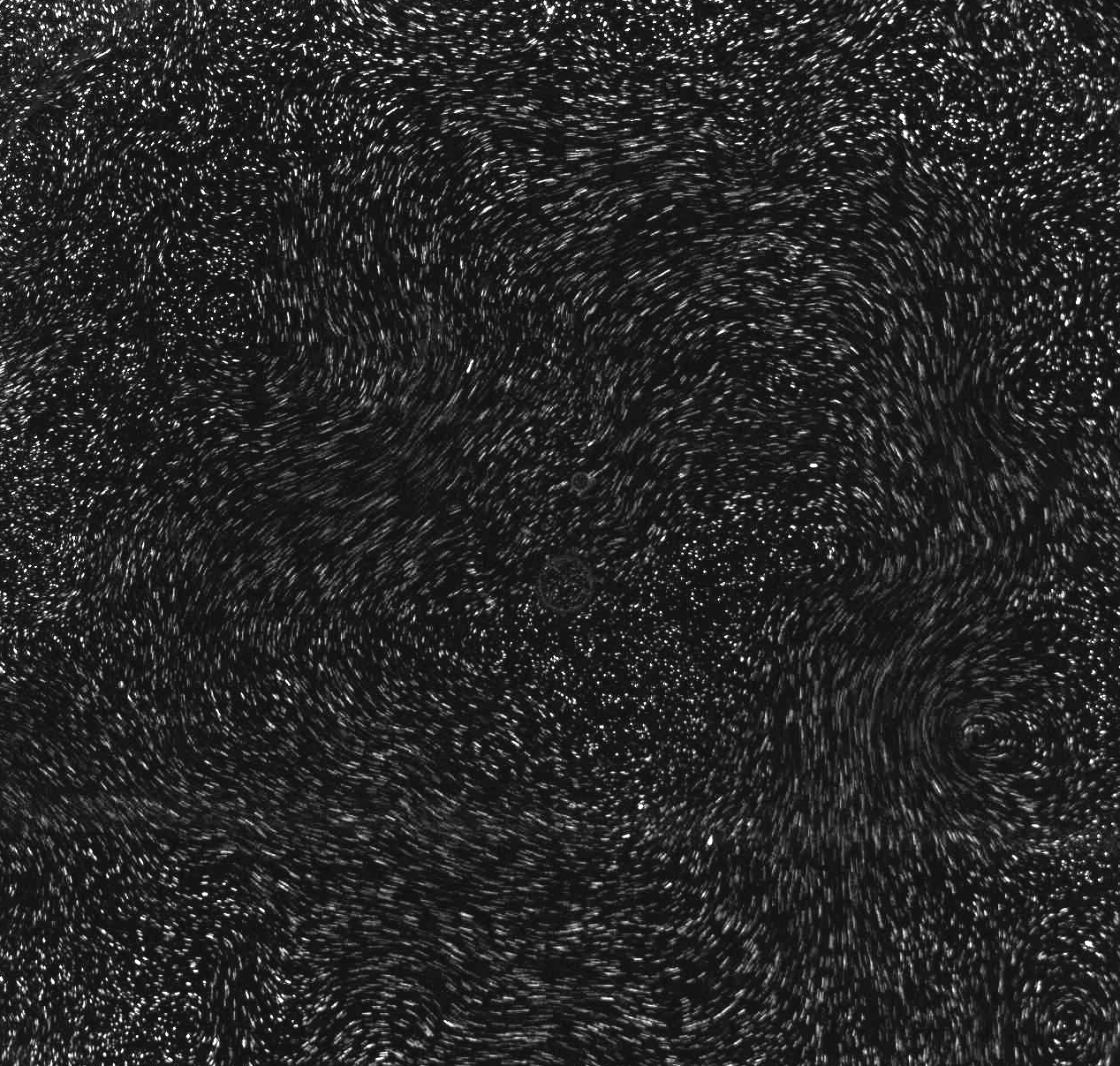}
	\caption{Streaks image ($1024\times1024$ pixels) from a longitudinal libration experiment in a cylindrical cavity. See section \ref{experiment} for details}
	\label{fig:expImage}
\end{figure}

We aim at inferring the displacement and azimuth from a streak image of $N \times N$ pixels, typically as illustrated in Figure \ref{fig:expImage}.

Similarly to how PIV is applied to a pair of images, we split the image into ($n \times n$, with n=48 in this study) sub-windows with an overlap ($50\%$ in our case). Assuming the velocity is sufficiently uniform over the ($n \times n$), we aim at determining the mean length, the displacement over the exposure time, and azimuth for each sub-window. 

A closer look at the image shows that the number of streaks, their intensity and their width, vary within each sub-window. In addition, in many cases, the streaks leave the ($n \times n$) interrogation area. In order to make our CNN resilient to this variability, we generate train and validation images of ($48 \times 48$ pixels) with a random number of streaks varying between 2 and 10, a random color intensity between 90 and 255 and a random thickness between 2 and 4 pixels. The center of each streak is chosen randomly over the entire sub-window, such that streaks are allowed to go outside the image.

To generate train and validation images, we sampled displacement, $\Delta$, and asimuth, $\phi$, from uniform distributions
\begin{equation}
\Delta \sim U(1, \Delta_\textnormal{max}) 
\end{equation}
and
\begin{equation}
\phi \sim U(-\pi/2+\delta, \pi/2),
\end{equation}
respectively. 

Here, $\phi = 0$ corresponds to the positive $x$-axis and increases clockwise. 
Note that limits for $\phi$ cover only half of the circle minus $\delta = 5^{\circ}$ on one end to mitigate ambiguity with respect to the direction. Once streaks are generated, the intensity within each streak intensity is perturbed and the image is blurred with a Gaussian kernel of $3\times 3$ pixels. In this study, the maximum displacement, $\Delta_\textnormal{max}$, of a particle within an image is assumed to be half of the chosen window size, that is $r_\textnormal{max} = 24$ pixels. 

\begin{figure}
\includegraphics[width=\textwidth]{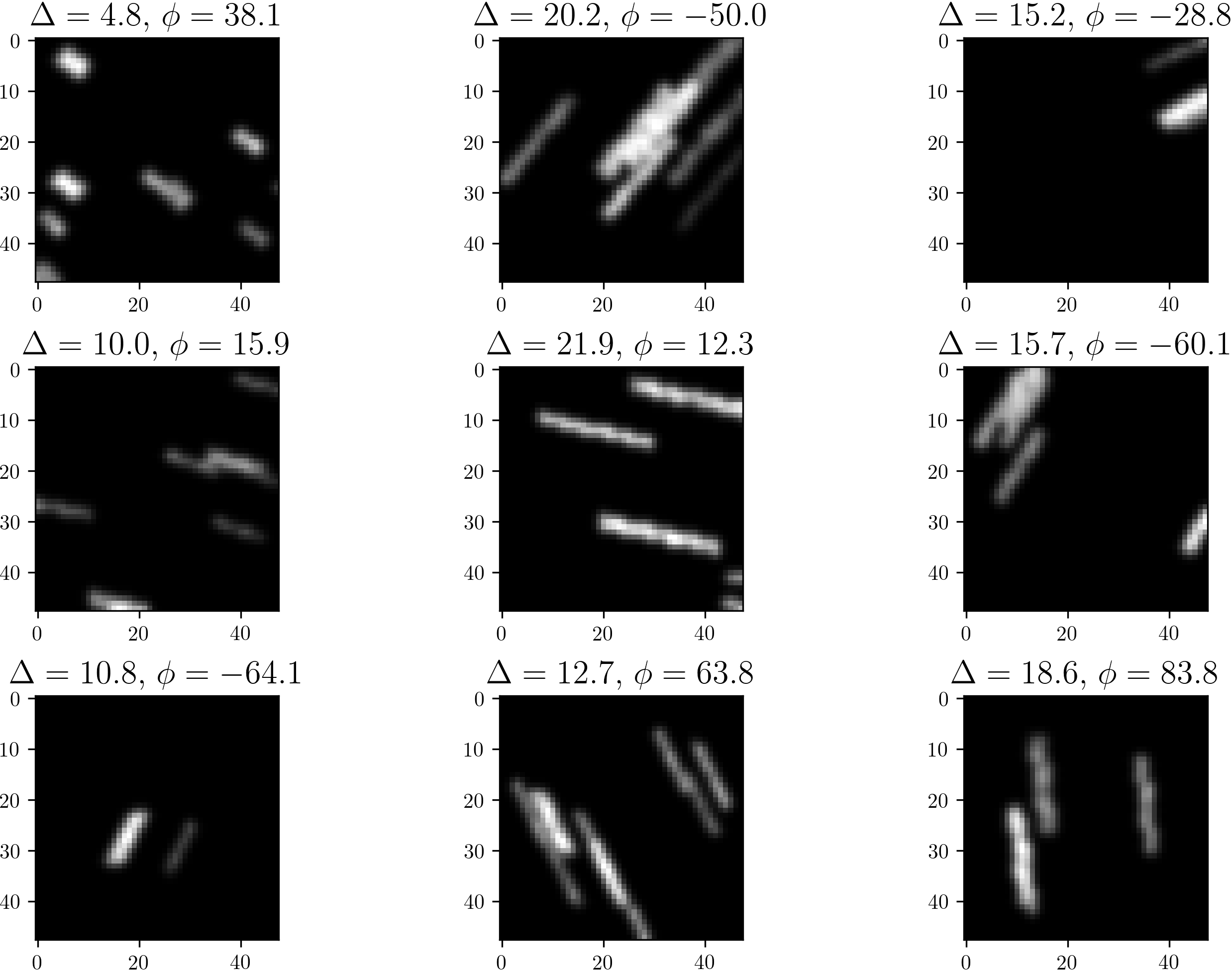}
\caption{Subset of generated streak images used for training and validation with the corresponding displacements (in pixels) and azimuth angles (in degrees) given in the titles.}
\label{fig:fig1}
\end{figure}

Figure \ref{fig:fig1} shows a selection of streak images generated using the stated procedure. The outlined procedure enables generation of millions of images in just a few minutes on a regular PC. 

\subsection{Network architecture and implementation}

\begin{figure}
\includegraphics[width=\textwidth]{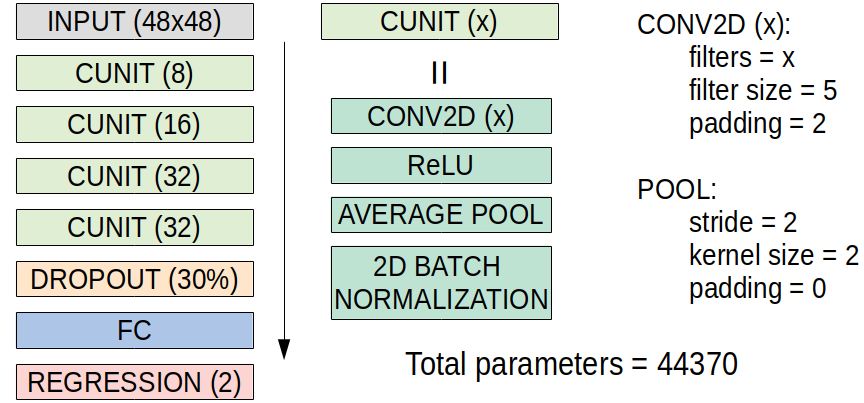}
\caption{Architecture of the used network. Left: sequence of layers and corresponding parameters in the brackets. Each CUNIT layer consists of four actual layers listed in the middle column. Relevant layer parameters are given on the right. Total number of learnable parameters in the network is 44370.}
\label{fig:figcnn}
\end{figure}

We aim at building a regression CNN that, given a single window image, outputs two real numbers, corresponding to the displacement and angle. To this end, we have created a network with the architecture shown in Figure \ref{fig:figcnn}. As can be seen, four convolutional units form the core of the network, each consisting of a convolutional layer, activation layer (in this case, rectified linear unit), average pooling layer used to downsample the input followed by a batch normalization layer. An increasing depth of the convolutional layers is a common choice aimed at giving a network ability to learn more complex features from the input. 

To prevent overfitting and improve training speed, we used batch normalization at the end of each convolution unit \cite{ioffe2015batch} and a dropout layer with the 30\% drop fraction \cite{srivastava2014dropout}. Additionally, following common practices, the outputs are standartized and centered. The input grayscale images have a single eight bits channel and are normalized by 255 to stay in the $[0, 1]$ range.

Whereas there may exist more efficient architectures, we found that for our problem increasing depth of the network has not improved overall performance significantly, nor did increasing and/or decreasing filter sizes. Additionally, we found that, in contrast to majority of the reported CNN architectures, using the average pooling instead of the max pooling results in a slightly higher accuracy.

To increase accuracy and robustness of the prediction, we built an ensemble consisting of ten networks \cite{hinton2015distilling}. All networks have the same architecture and parameters, but were trained using different randomly chosen initial weights \cite{glorot2010understanding, he2015delving} and random shuffling to form a unique sequence of batches supplied to an optimizer. The outputs of all ensemble members are averaged to get a the final prediction and its variance, thereby providing a proxy for the uncertainty of a prediction.

Finally, the network was implemented using the PyTorch library \cite{paszke2017automatic}. The complete implementation can be found in GitHub using the link below.

\section{Results}

\subsection{Network training}

We generated one million images (e.g., Figure \ref{fig:fig1}) of which 75\% and 25\% were used for training and validation, respectively. Each network in the ensemble has been trained for 100 epochs. The ADAM optimizer with an initial learning rate value of 1e-3 was used. The learning rate was halved if no sufficient decrease in the validation loss was observed for the past 10 epochs. To make the training more resistant to potential outliers (e.g. imagine a situation when all streaks leave the window), we chose to minimize the Huber loss \cite{huber1973robust} rather than conventional least-squares loss. Huber loss represent a hybrid $L_2-L_1$ distribution with a high tolerance to the presence of long tails in the original distribution. 
No pathologies between training and validation losses were observed during the training, specifically the validation loss remained a bit higher than the training loss, indicating that the network learns some generic features of the dataset. Training of each network took $\approx 90$ minutes on a single CPU-GPU system. This modest time suggests that training networks for window sizes larger that the one adopted here is feasible. 

\begin{figure}
\includegraphics[width=\textwidth]{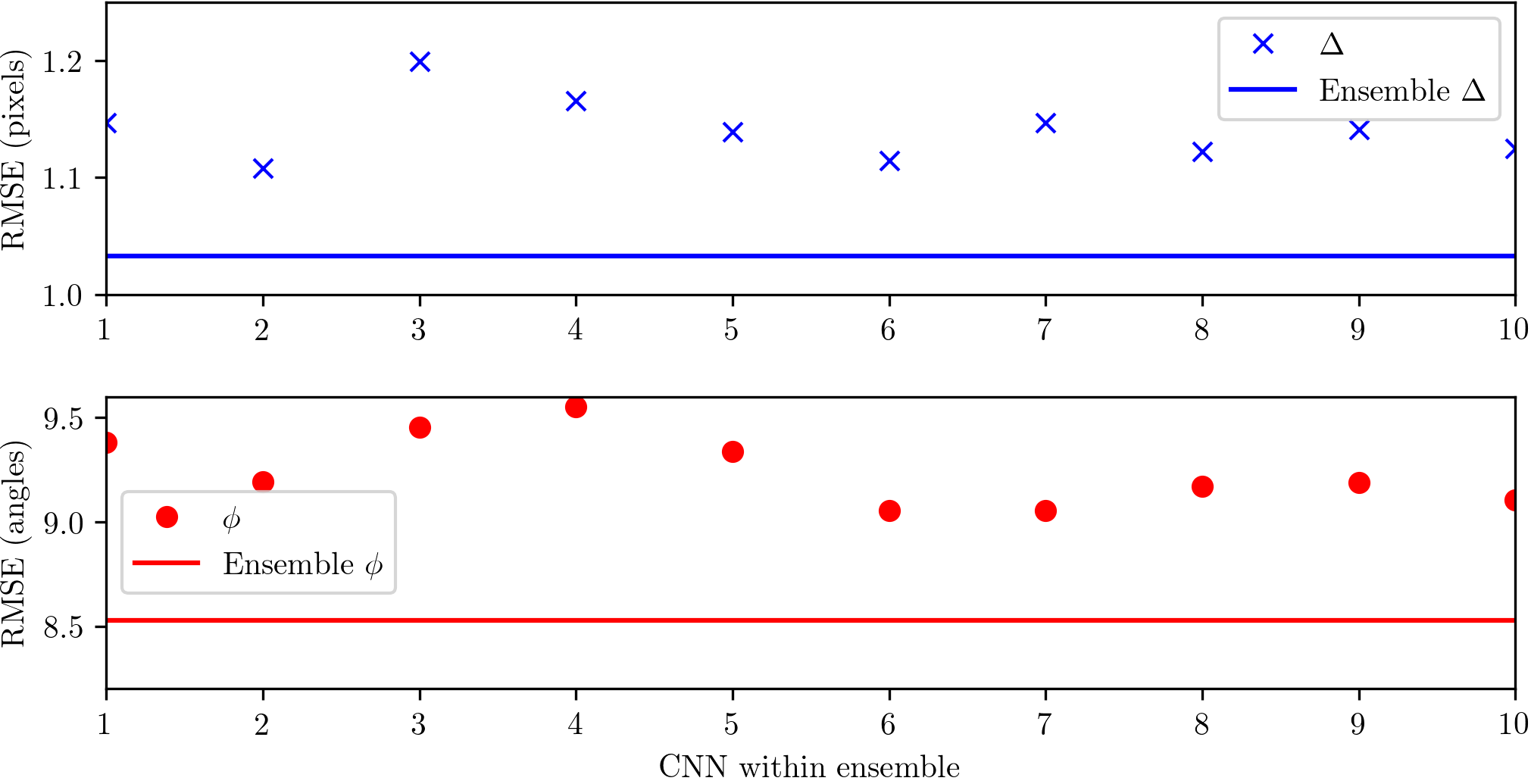}
\includegraphics[width=\textwidth]{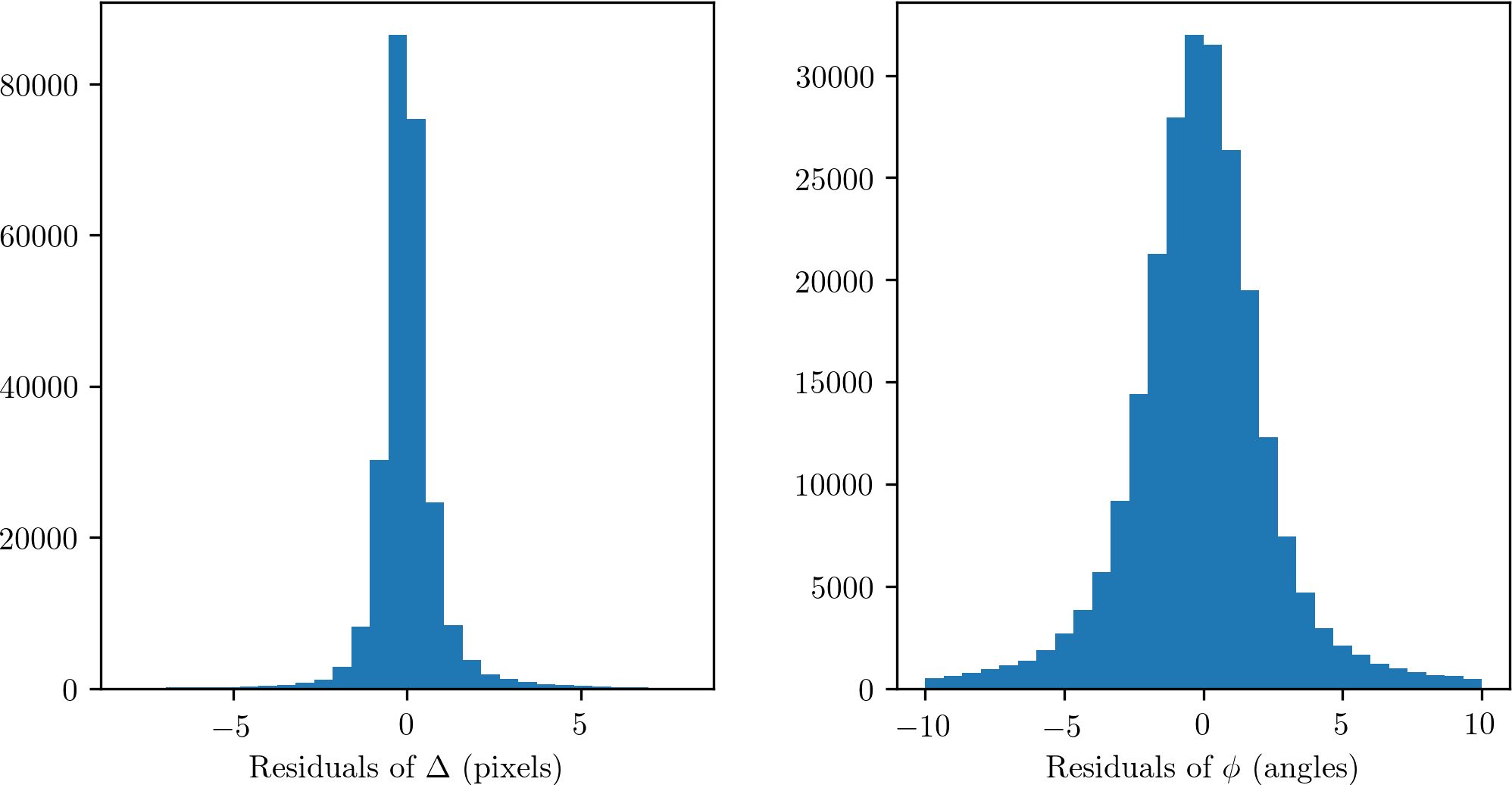}
\caption{Top: RMSE values of individual networks (markers) and ensemble (line) for displacement $\Delta$ and angle $\phi$ calculated on the validation set. Bottom: histograms of the residuals for the displacement and angle.}
\label{fig:fig3}
\end{figure}

Figure \ref{fig:fig3} shows RMSE values for all networks seperately as well as the RMSE of the ensemble. It is evident that the ensemble performs up to ~20\% better than a single network. CNN ensemble achieves the accuracy of 1.05 pixels for the displacement and $8.5^{\circ}$ for the angle predictions, respectively. The latter may seem like a large error, however one should realize that having a $48 \times 48$ pixels window inevitably leads to limitations in recognizing small rotations, especially when streaks are short. 

\begin{figure}
\includegraphics[width=\textwidth]{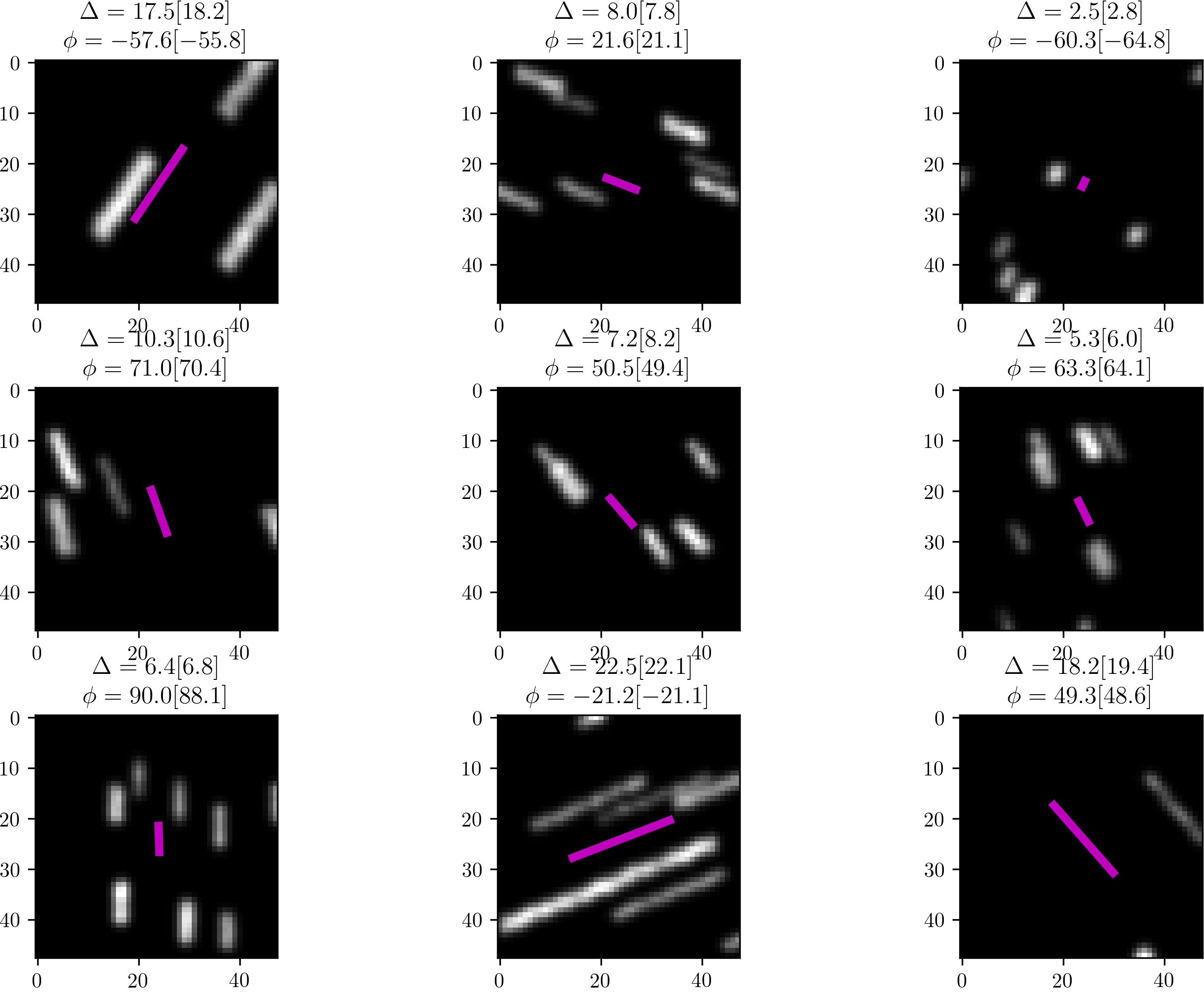}
\caption{Images with the predicted displacement and angle depicted as magenta lines. The true values are given in titles and predicted values in brackets.}
\label{fig:fig4}
\end{figure}

\begin{figure}
\includegraphics[width=\textwidth]{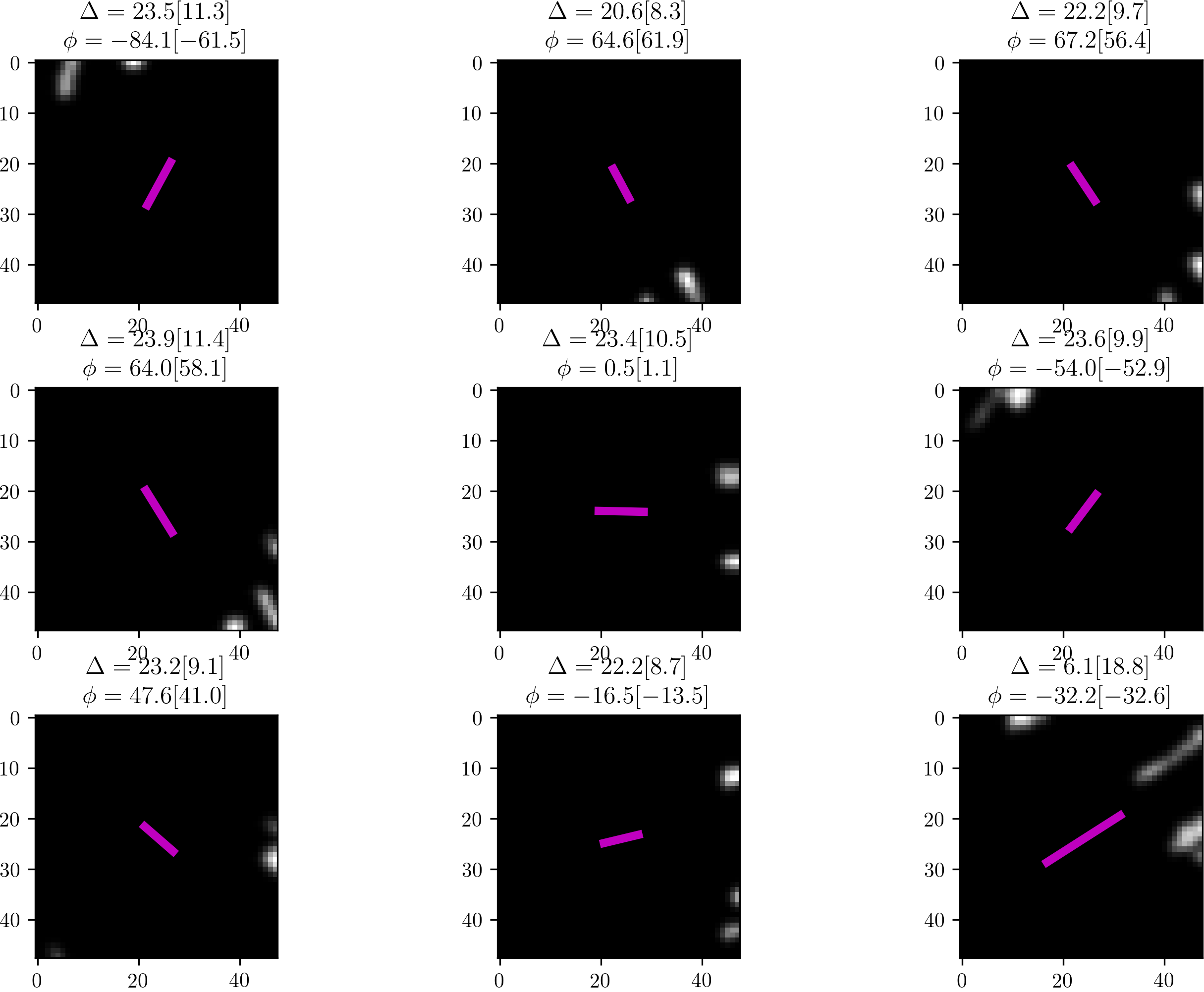}
\caption{Same as previous, but showing nine worst prediction in terms of displacement.}
\label{fig:fig5}
\end{figure}

Figure \ref{fig:fig4} shows a random selection of validation images with the corresponding ensemble CNN predictions and true values. We see that the ensemble CNN has learned to make accurate prediction even for complex situations when some streaks overlap, join or leave the window. Additionally, Figure \ref{fig:fig5} shows nine images which produced the worst predictions of the displacement. Clearly, most of these cases are associated with situations where originally longer streaks have all been displaced outside the window or multiple streaks coincidentaly joined and formed what appears to be a longer streak. Note that such situations do not represent a failure of the CNN, but rather limitation of the window-based approach adopted in this work. Similar situations are likely to occur in experimental images and ufortunately they leave little chance for any sliding window approach to make an accurate prediction. We will discuss possible ways to mitigate this in the Conclusions section. It is worth to note that despite inaccurate displacement prediction, the angle is predicted accurately in most of these cases. 

\subsection{Application to a numerical simulation}

\begin{figure}
\includegraphics[width=\textwidth]{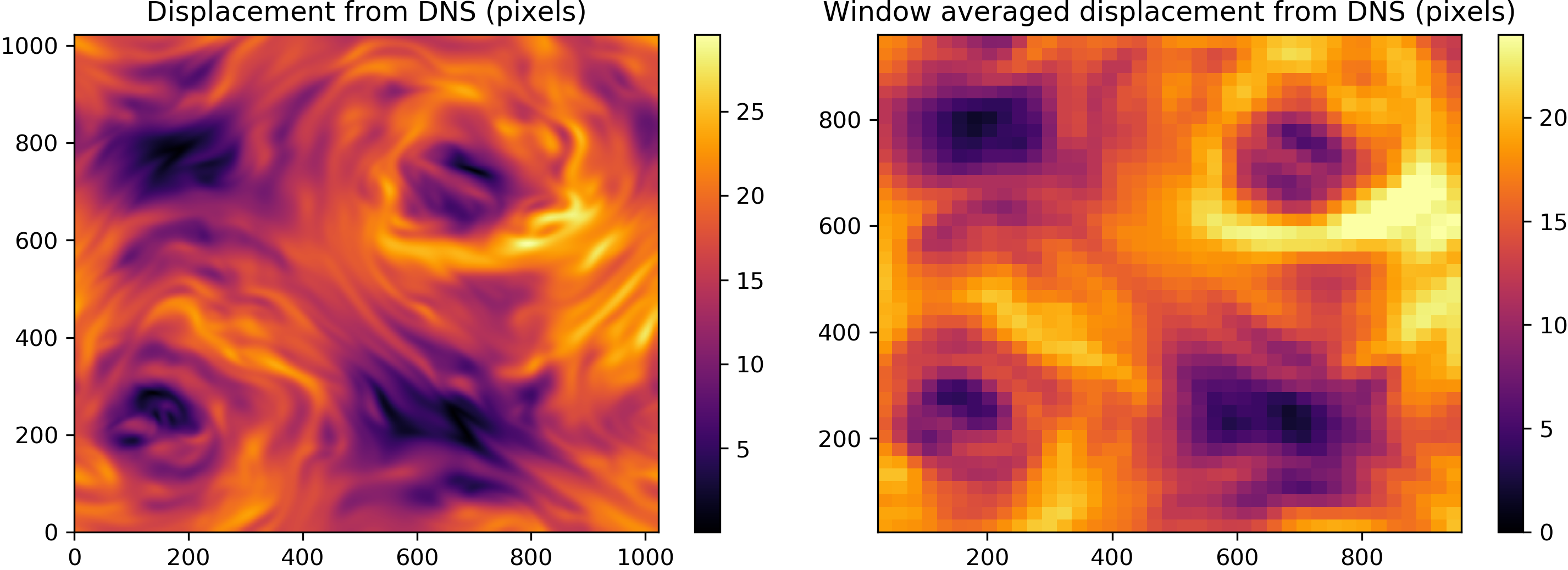}
\caption{Original (left) and window averaged (right) displacement maps of the DNS \cite{plumley2016effects} used for validation.}
\label{fig:fig6}
\end{figure}

After confirming the excellent performance of the network on the validation set, we turn to a discrete numerical simulation (3D-DNS) of a complex thermal convection turbulent flow presented recently by \cite{plumley2016effects}. In their paper, the authors simulate rapidly rotating thermal convection in a square box, heated from below and cooled from the top with a vertical gravitational field pointing downward and a vertical axis of rotation. To validate our CNN algorithm, we used a simulation obtained for $E=10^{-7}$, where $E$ is the Ekman number representing the ratio of viscous to rotational effect, and a Raleigh number that is 90 times supercritical. Each velocity field is $192^3$. For the purpose of this study, we extracted the horizontal component of the velocity in a horizontal plane a mid-depth. Due to the limited size of the calculation each velocity component has been interpolate on a $1024\times 1024$ pixels grid. 
Figure \ref{fig:fig6}-left shows the displacement map obtained from the full DNS for an arbitrary exposure time. Since the CNN will act on ($48\times48$) interrogation windows with the $50\%$ overlap, we apply a sliding window averaging with ($48\times48$) and $50\%$ overlap (Figure \ref{fig:fig6}-right). The window averaged displacement is the best reconstruction we can possibly achieve with a $48 \times 48$ pixels window and a 50\% overlap used in this study, thus all later results will be compared against it.

\begin{figure}
\includegraphics[width=\textwidth]{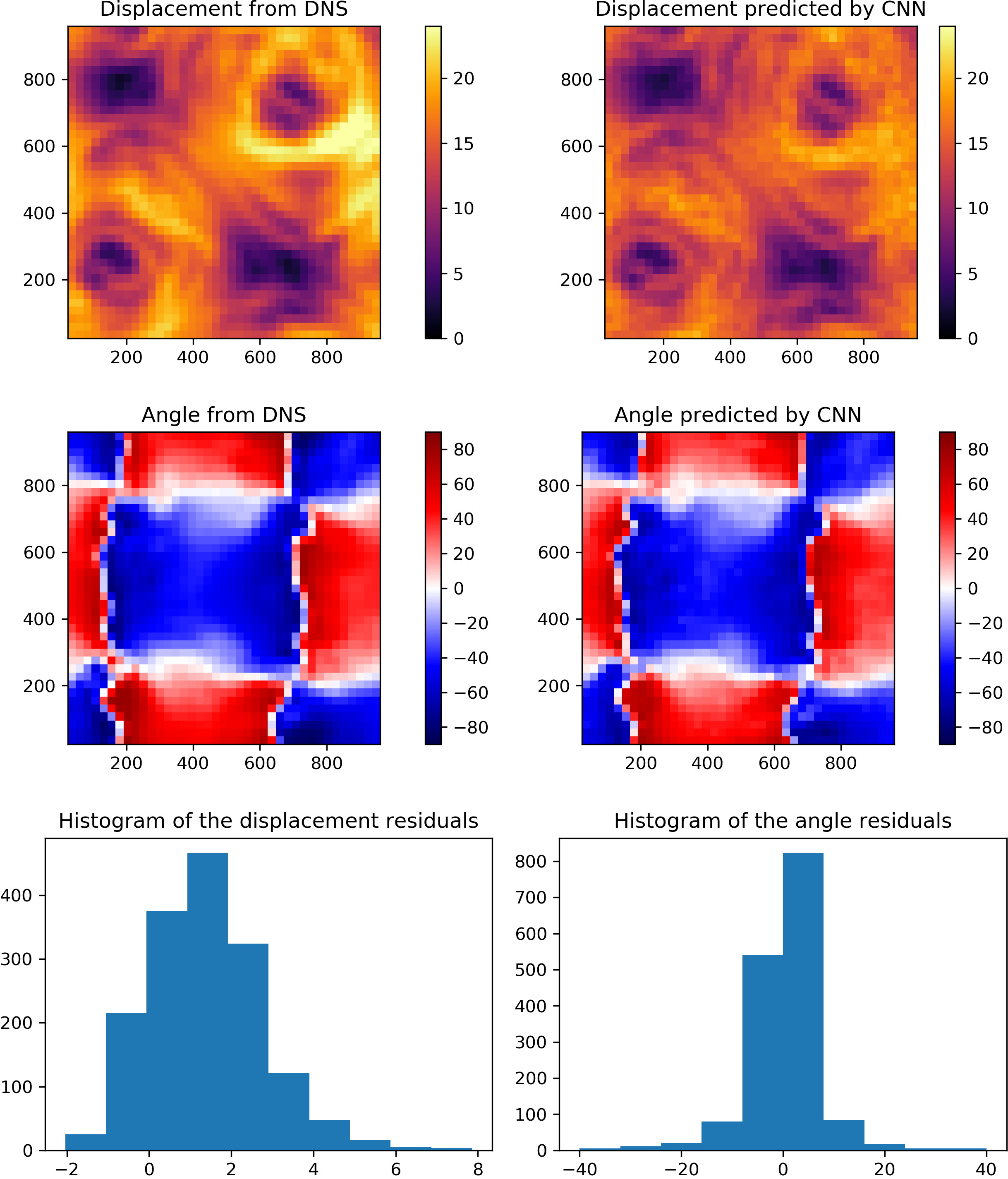}
\caption{Original (left) and window averaged (right) displacement maps of the DNS \cite{plumley2016effects} used for validation.}
\label{fig:fig7}
\end{figure}

To reduce subjectivity, we generated 30 randomly seeded streak images using the original DNS velocity (Figure \ref{fig:fig6}, left) and varying streak density, intensity and thickness as formerly stated. The trained CNN ensemble was applied to all images and the resulting averaged reconstruction of both displacement and angle are shown in Figure \ref{fig:fig7}. We see that the reconstructed images recover the general structure of the flow very well. The only undesired effect is that the ensemble CNN struggles to correctly predict large displacements. When the displacement reaches half of the window size, CNN systematically predicts smaller displacements. The reason for this is not a deficiency of the CNN, but rather the fact that for displacements as large as half the window size it becomes very likely that most of the streaks go outside the window, thus a smaller displacement is a rational prediction. To solve this problem, one could apply a larger window (for instance, $64\times 64$ pixels), although this will reduce the overall resolution and may render constant velocity assumption less accurate. In real experimental settings, it may be advantageous to apply several window sizes and either choose one or combine obtained results, depending on a particular experimental setup.

\begin{figure}
\includegraphics[width=\textwidth]{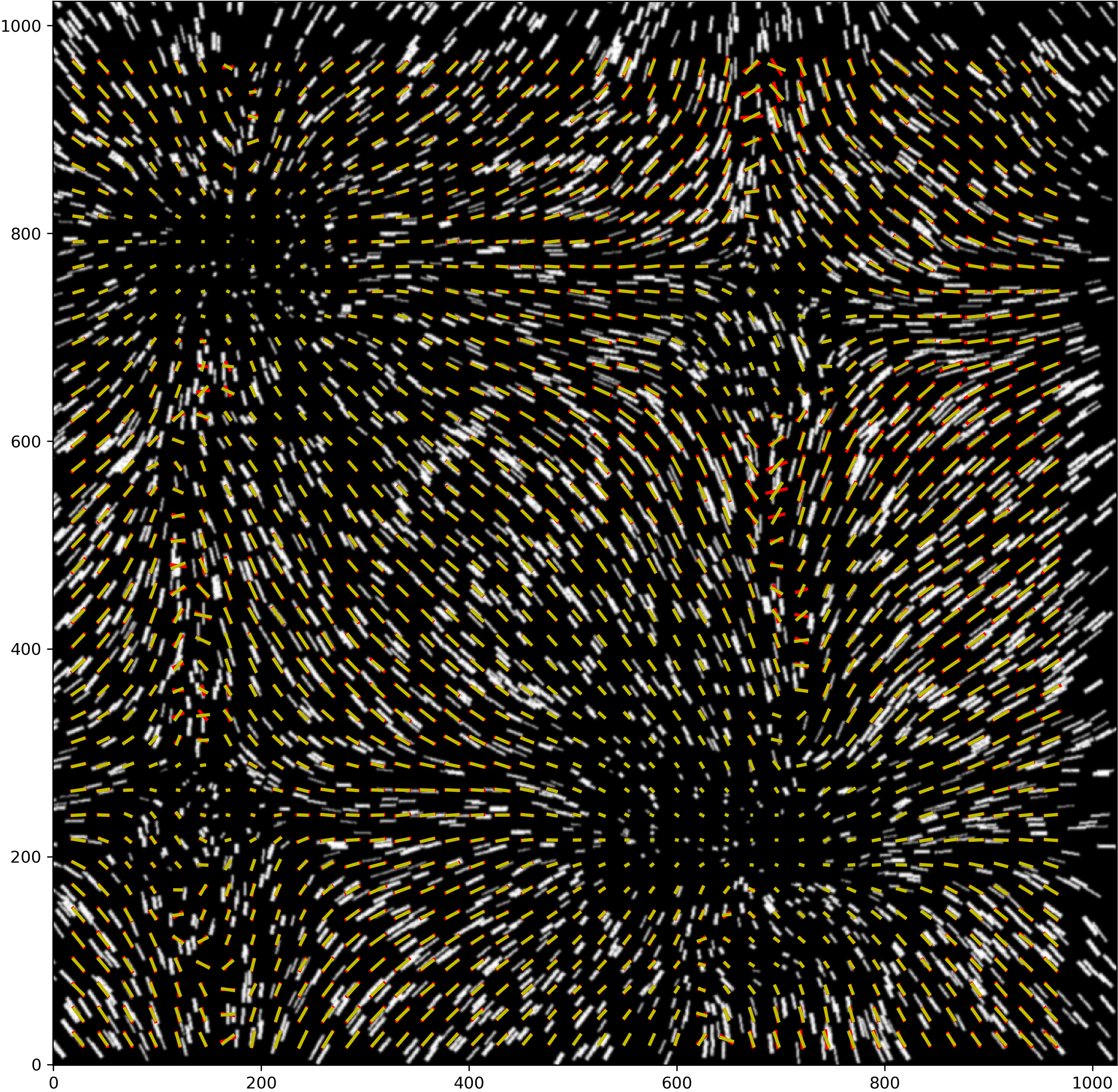}
\caption{Streak image generated from the DNS with true and predicted displacement and angle shown as red and yellow lines, respectively.}
\label{fig:fig8}
\end{figure}

Finally, Figure \ref{fig:fig8} shows one of the streak images generated from DNS with the true and predicted streaks for each window position. Note that while the whole velocity field is used to generate the background streak images, the overlying red streaks are reconstructed at the location of the CNN grid point using the averaged displacement and angle maps (Figure \ref{fig:fig6}) Generally, ensemble CNN performs well. There is, however, accuracy reduction in regions of large displacements for reasons described above and across areas of vertical shearing, where angle predictions become locally less accurate. Interestingly, these vertical shear regions also lead to abnormal streaks orientation (close to horizontal) when using the window averaged field values (red streak in Figure \ref{fig:fig6}).

\subsection{Application to the flow driven by longitudinal libration in laboratory experiment}\label{experiment}
\begin{figure}
	\includegraphics[width=\textwidth]{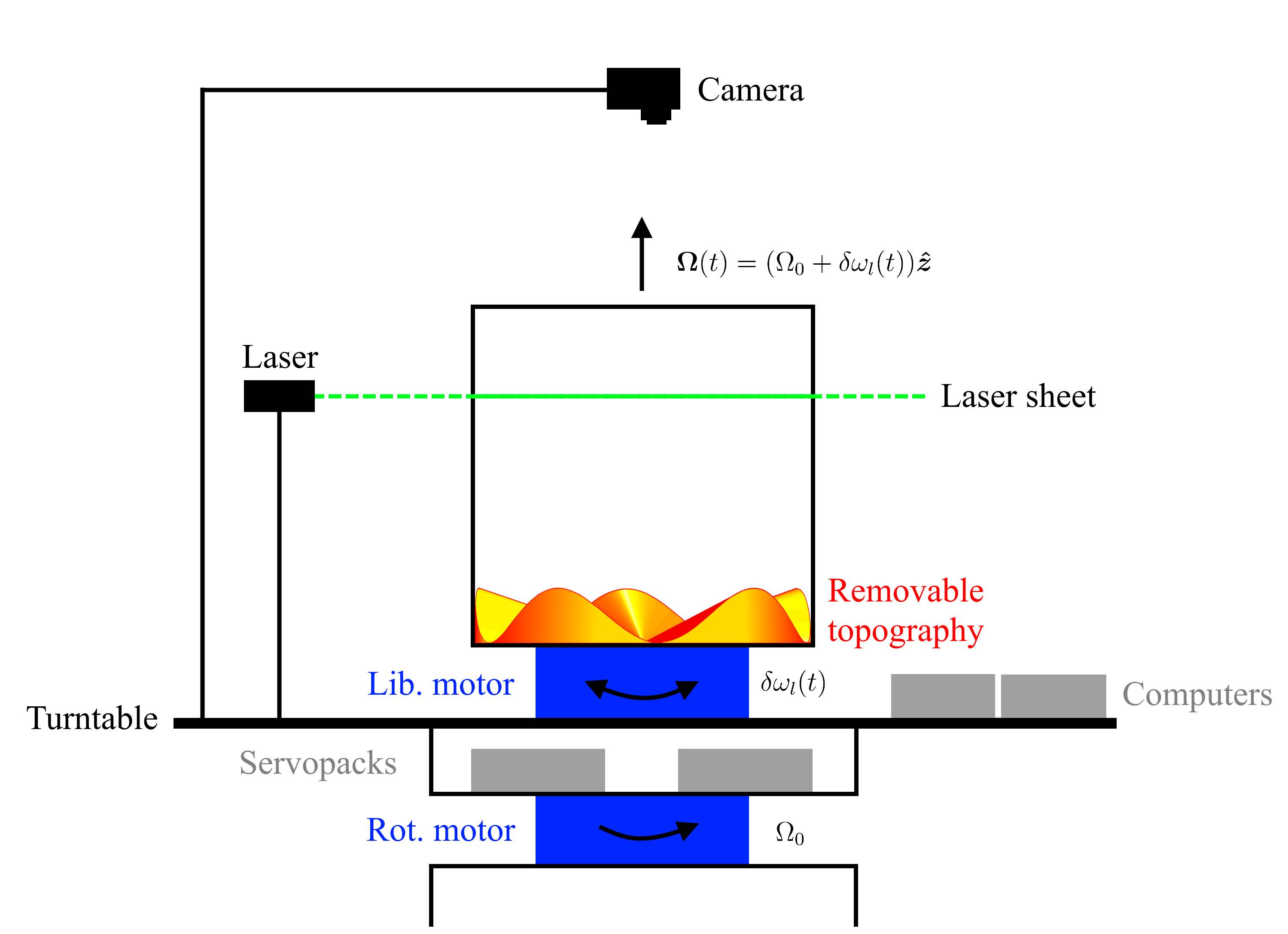}
	\caption{Experimental setup of the longitudinal libration experiment \cite{MeisterMSC}.}
	\label{fig:MeisterMSC}
\end{figure}

Finally, we apply our algorithm to the laboratory experiment depicted in figure \ref{fig:MeisterMSC}. The apparatus consists of a straight cylinder filled with water set in rotation on a turntable at 1Hz. The cavity is 286mm high with a radius of 140mm. The bottom lead is covered with topography made of blocks ($64mm\times64mm\times 8mm$). A second motor is used to oscillate the container on the turntable, resulting in the so-called longitudinal libration. For the purpose of this experimental validation we have set the libration frequency to 1Hz and the amplitude to $34^{\circ}$. 

To obtain the streak images, we use a 1W continuous diode laser to illuminate a horizontal plane of the fluid seeded with fluorescent particles. We record movies of the particles motion at a fixed exposure time of 21ms with a sampling frequency of 30 images per second. We extracted a frame corresponding to a maximum apparent amplitude of the swirling flow (\ref{fig:expImage}).
     
A pre-processing step was necessary for real images to filter out the CCD matrix noise. We achieved this by a simple thresholding of an image at intensity of 50. After that, image was split into overlapping $48\times 48$ pixels windows, resulting in a total of 2400 window images. The ensemble CNN was then applied to predict displacements and angles as well as their standard deviations with the latter serving to be a proxy for the output uncertainties. 

\begin{figure}
\includegraphics[width=\textwidth]{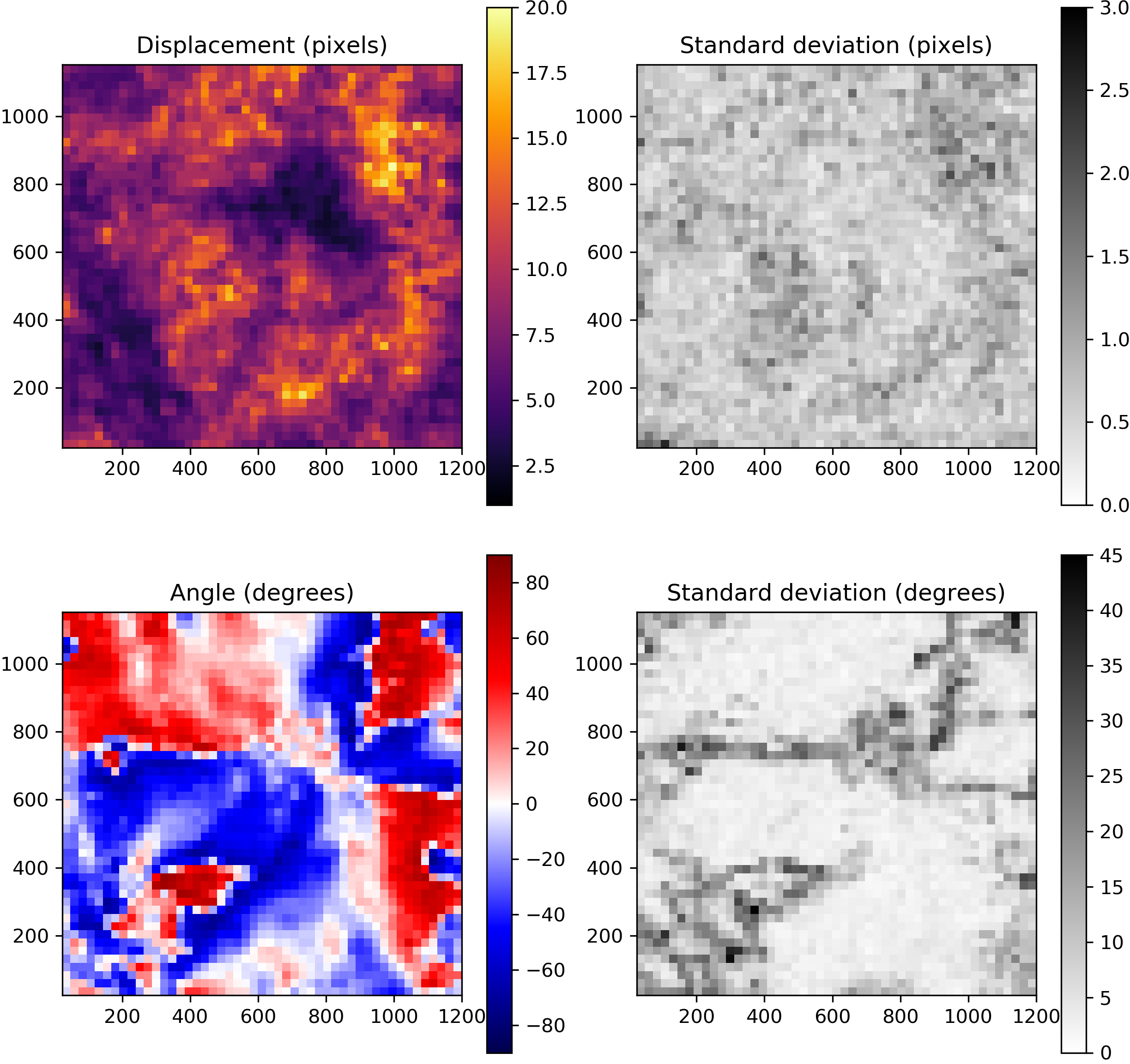}
\caption{Predicted displacement and angle (left) along with their standard deviations (right) inferred from the ensemble CNN for an experimental image shown in Figure \ref{fig:fig10}.}
\label{fig:fig9a}
\end{figure}

\begin{figure}
\includegraphics[width=\textwidth]{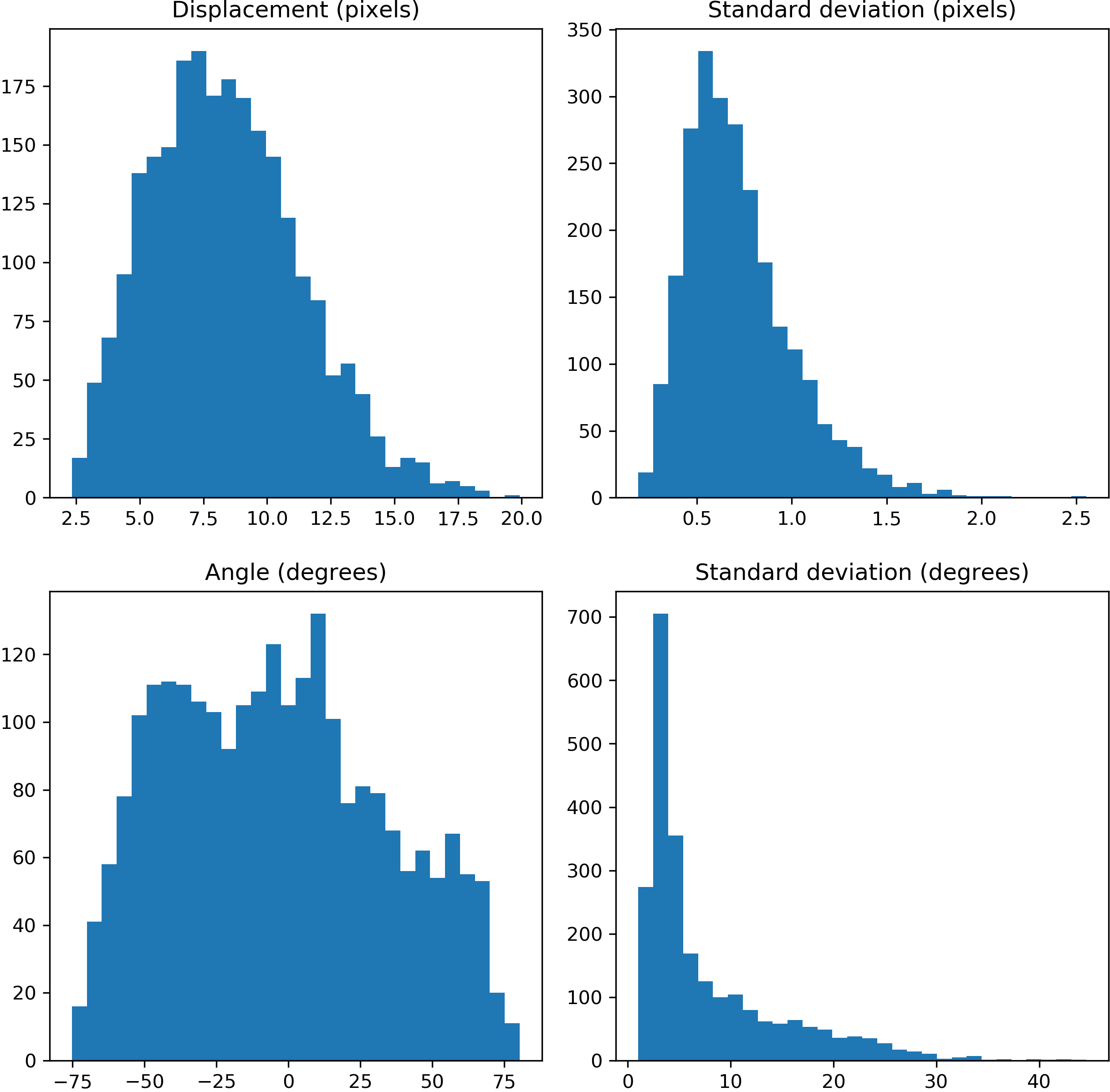}
\caption{Same as above, but in a form of histograms.}
\label{fig:fig9b}
\end{figure}

\begin{figure}
\includegraphics[width=\textwidth]{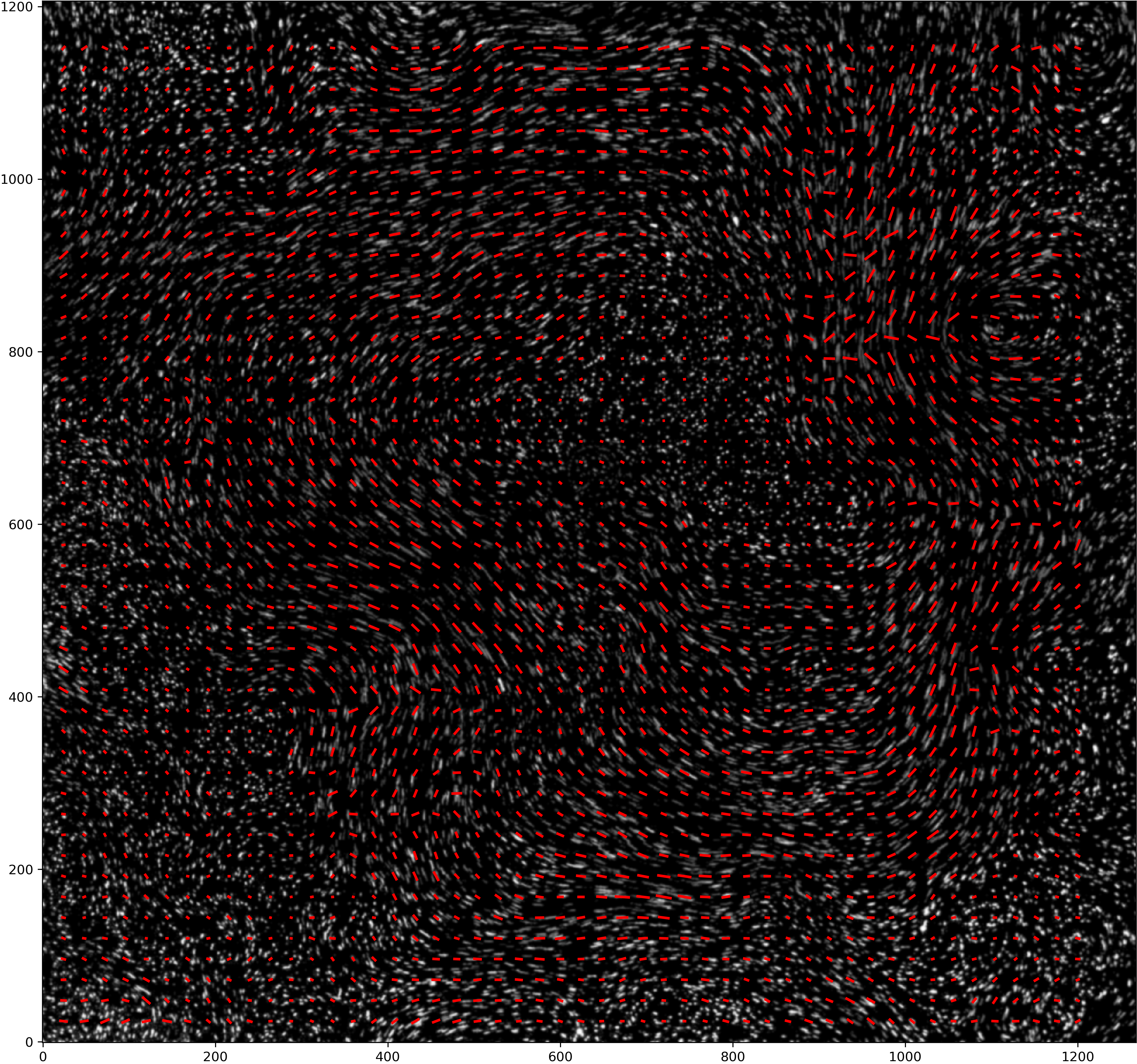}
\caption{Streak image taken during an experiment overlain by the predicted displacement and angle shown as red lines.}
\label{fig:fig10}
\end{figure}

Figures \ref{fig:fig9a}-\ref{fig:fig9b} show the predictions in the form of maps and histograms, respectively. Figure \ref{fig:fig10} shows the actual experimental image we used and window-centered predictions drawn as lines. First of all, we see that experimental flow exhibits rather involved behaviour, with regions where streaks show non-linear behvaiour of the flow within a window, which violates our initial assumptions. Nonetheless, the ensemble CNN produces coherent and spatially correlated predictions (Figure\ref{fig:fig9a}) It is also interesting to analyse the standard deviation maps (Figure \ref{fig:fig9b}). Large absolute errors in the displacement map appear to be correlated with regions of fast flow. On the other hand, large uncertainties in the angle occur around regions with vertical velocity shear, as observed in DNS synthetic images, specifically where velocity seems to flip the direction. 

In terms of performance, ensemble CNN attains a speed of seven full images (each totaling to 2400 windows) per second on an NVIDIA GoForce GTX 1070Ti powered PC. This level of performance allows us to comfortably process long sequences of images taken during experiment, which we use to study evolution of the kinetic energy. Using a more modern GPU will likely also enable real-time predictions at rates of a few dozens of images per second, which could be more relevant for industrial applications.

\section{Conclusions}

We presented an open-source ensemble CNN aimed at quantitative analysis of complex experimental flows using streak images. The presented approach can be applied in situations when classical PIV is inaccurate or not possible due to experimental setup or hardware limitations. The advantage of the presented approach is the ease of training/validation set generation as well as recovery of the prediction uncertainty through application of the ensemble method. We foresee that the approach may need an experiment-dependent tailoring to achieve the best performance. According to our experience, this only takes a single day on an average PC equipped with a modern GPU and should not represent a practical limitation.

The potential extensions of the presented approach may include the use of multiple window sizes applied to the same image, more complex functional parameterizations of the displacement and angle fields within a window (e.g., polynomial). Finally, a completely different approach based on an image segmentation and object localization and/or tracking can be applied to extract information about individual streaks and their evolution in time. We expect to investigate these directions in details in our future studies.

\begin{acknowledgements}
We thank authors of PyTorch, mathplotlib and h5py libraries for making them available, Meredith Plumley for sharing her DNS data, Adrian Tasistro-Hart for a number of insightful discussions on CNNs. The Jupyter Python notebooks of all programs used in this study can be found here \url{https://github.com/agrayver/streakcnn}.
\end{acknowledgements}

% References
%\bibliography{refs}


\end{document}